\documentclass{article}

\usepackage{amssymb,amsfonts,amsmath}
\usepackage{cite,enumerate,float,indentfirst}
\usepackage{color}

\def\be{\begin{eqnarray}}
\def\ee{\end{eqnarray}}
\def\nn{\nonumber}

\def\p{\partial}
\def\tr{{\rm tr}\,}
\def\Tr{{\rm Tr}\,}

\definecolor{red}{rgb}{1,0,0}
\definecolor{orange}{rgb}{1,0.5,0}
\definecolor{violet}{rgb}{0.7,0,1}

\hoffset= - 1.0in         % switch off for draft style
\begin{document}

\title{\vspace{-.5cm}{\Large {\bf  Hook variables: cut-and-join operators and $\tau$-functions}\vspace{.05cm}}
\author{
{\bf A.Mironov$^{a,b,c}$}\footnote{mironov@lpi.ru; mironov@itep.ru}\ \ and
\ {\bf A.Morozov$^{d,b,c}$}\thanks{morozov@itep.ru}}
\date{ }
}

\maketitle

\vspace{-5cm}

\begin{center}
\hfill FIAN/TD-14/19\\
\hfill IITP/TH-20/19\\
\hfill ITEP/TH-34/19\\
\hfill MIPT-TH-18/19
\end{center}

\vspace{3cm}

\begin{center}
$^a$ {\small {\it Lebedev Physics Institute, Moscow 119991, Russia}}\\
$^b$ {\small {\it ITEP, Moscow 117218, Russia}}\\
$^c$ {\small {\it Institute for Information Transmission Problems, Moscow 127994, Russia}}\\
$^d$ {\small {\it MIPT, Dolgoprudny, 141701, Russia}}
\end{center}

\vspace{.5cm}

\begin{abstract}
Young diagrams can be parameterized with the help of hook variables,
which is well known, but never studied in big detail.
We demonstrate that this is the most adequate parameterization
for many physical applications:
from the Schur functions, conventional, skew and shifted,
which all satisfy their own kinds of determinant formulas
in these coordinates,
to KP/Toda integrability and related basis of cut-and-join $\hat W$-operators,
which are both actually expressed through the single-hook diagrams. In particular, we discuss a new type of multi-component KP $\tau$-functions, Matisse $\tau$-functions.
We also demonstrate that the Casimir operators, which are responsible for integrability, are single-hook, with the popular basis of ``completed cycles"
being distinguished by especially simple coefficients in the corresponding expansion.
The Casimir operators also generate the $q=t$ Ruijsenaars Hamiltonians.
However, these properties are broken by the naive Macdonald deformation,
which is the reason for the loss of KP/Toda integrability and related structures
in $q$-$t$ matrix models.
\end{abstract}

\vspace{.0cm}

\section{Introduction}

Schur functions \cite{Mac} play increasingly important role in modern mathematical physics.
They are 
\begin{itemize}
\item characters of representations of the $GL_\infty$ group \cite{Weyl};
\item solutions to the Hirota bilinear equations \cite{DJKM,Sat};
\item eigenfunctions of the Calogero and $W$-operators \cite{Cal,MMN}.
\item They form a basis of functions, which have factorized Gaussian matrix averages,
and appear in $<{\rm character}> = character$ relations \cite{OrMM}.
\item They are starting point for various triangular deformations,
which lead to Jack, Macdonald and Kerov functions \cite{Mac,Kerov,MMKerov}.
\end{itemize}
At the same time, they still lack a clear first-principle definition,
which puts the entire field on a somewhat shaky ground.
The most popular interpretation is related to free fermion representations \cite{DJKM,JM},
which are peculiar to $k=1$ Kac-Moody algebras,
which causes difficulties with applications to other $k$,
where Wakimoto-like representations in terms of free bosons are needed \cite{Wak}.
Instead, the free fermion formalism provides nice determinant formulas
for the Schur functions \cite{JC,DJKM},
which are well suited for applications to KP/Toda integrability problems \cite{DJKM},
while, in the free boson case, determinants are changed for more general combinations
of Riemann theta-functions, and integrability is not immediately transparent.

In this paper, we discuss determinant representations for all kinds of Schur and $\tau$- functions
in the hook (Frobenius) variables and related topics: the story is well known to experts, but have not received
enough publicity so far.
Our hope, however, is that these ``weaker" determinant formulas may appear more
``fundamental" and more suitable for further generalizations.
In particular, we demonstrate that they are applicable to the {\it shifted}
Schur functions \cite{OO}, which are also solutions to the Hirota integrable equations.
The hook variables are also related to the skew characters and show up not-quite-expectedly
in the differential-expansion coefficients for twist and double-braid knots \cite{Mor16}.
They also provide non-trivial solutions to the Pl\"ucker relations
and thus provide new types of KP $\tau$-functions \cite{ON},
somewhat closer to the partition functions of the rainbow tensor models \cite{IMMbr,IMM,IMMint}.
Last but not least, the Casimir operators are essentially single-hook, and they generate the ordinary $\tau$-functions, this is what makes these latter distinguished among the Hurwitz $\tau$-functions, which do not yet possess any clear description in terms of Hirota-like equations.

This broad variety of applications makes hook representations for the Schur functions
an interesting subject, which should be further explored.
This paper is just the first step on this way.

The paper is organized as follows. Sections 2 and 3 provide a preliminary preparation for the main part of the text, we discuss determinant representations of the Schur functions (sec.2) and of the skew Schur functions (sec.3) in terms of the hook Schur functions. Section 4 contains a review of the relations between KP $\tau$-functions and linear and bilinear combinations of the Schur functions.
In section 5, we discuss $\tau$-functions that are multilinear combinations of the Schur functions $\chi_R$.
In section 6, we discuss determinant representations of the shifted Schur functions \cite{OO} in terms of the hook Schur and shifted Schur functions. We also notice that the shifted Schur functions satisfy the Hirota bilinear equations.
In sections 7 and 8, we discuss the generalized cut-and-join $\hat W$-operators \cite{GD,MMN} and their representation in terms of hooks \cite{MMrev}. In particular, we discuss that, at each level, there is only one (up to a linear combination of lower level operators) cut-and-join $\hat W$-operator that involves only single-hook Schur functions.
At last, in section 9, we explicitly construct a simple basis of these operators, the Casimir operators associated with the completed cycles, and, in section 10, we evaluate the generating function of these operators and its eigenvalue. Section 10 contains a summary of the results obtained in the paper.

\section{Hook variables and determinant formula}

We denote the Schur functions as $\chi_R$. Schur functions \cite{Mac,Fulton} depend on infinite sequence of time-variables $p_k$, $k =1,\ldots,\infty$, they are graded polynomials of $p_k$, and are symmetric functions of variables $x_i$ on the Miwa locus
\be
p_k^{(N)} = \tr_{N\times N}  X^k = \sum_{a=1}^N x_a^k
\ee
The Schur function can be defined through the Frobenius formula \cite{Frob}
\be
\chi_R\{p_k\}=\sum_\Delta{\psi_R(\Delta)\over z_\Delta}p_\Delta
\ee
Here, for the Young diagram $\Delta=\{\delta_1\geq \delta_2\geq\ldots\delta_{l_\Delta}>0\}
= \{1^{m_1},2^{m_2},\ldots\}$,
the symmetry factor is defined $z_\Delta:=\prod_i m_i! \cdot i^{m_i}$ and $p_\Delta$ is a monomial
$p_\Delta \equiv p_{\delta_1}p_{\delta_2}\ldots p_{\delta_{l }}$, while
$\psi_R(\Delta)$ is a character of the symmetric group $S_n$, $n=|R|$. It satisfies the orthogonality conditions
\be\label{oc}
\sum_\Delta{\psi_R(\Delta)\psi_{R'}(\Delta)\over z_\Delta}=\delta_{RR'},\nn\\
\sum_R{\psi_R(\Delta)\psi_{R}(\Delta')\over z_\Delta}=\delta_{\Delta\Delta'}
\ee
as any character does.

Equivalently, one can construct the Schur function as a finite-dimensional determinant \cite{JC}. To this end,
one defines a Schur function for symmetric Young diagram $R=[r]$
through the expansion
\be
\exp\left(\sum_k \frac{p_kz^k}{k}\right) = \sum_{n=0}^\infty z^n\chi_{[n]}\{p\}
\ee
Then, for arbitrary Young diagram $R=\{r_1\geq r_2 \geq \ldots \geq r_{l_R} >0\}$, \cite{Mac}
\be\label{Ss}
\chi_R=\det_{ij}\chi_{[r_i-i+j]}
\ee

The hook (Frobenius) variables parameterize the Young diagram $R$ in a somewhat different way, $R=(\vec\alpha|\vec\beta)$ with $\alpha_1>\alpha_2>\ldots>\alpha_l>0$, $\beta_1>\beta_2>\ldots>\beta_l>0$:

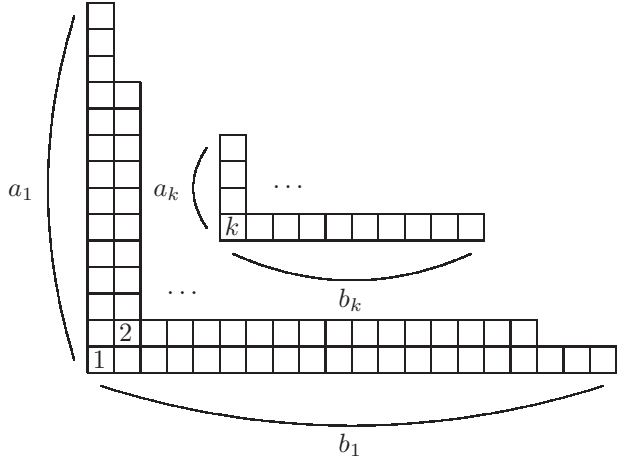
\begin{picture}(300,220)(-90,-55)

\put(0,0){\line(1,0){200}}
\put(10,10){\line(1,0){190}}
\put(20,20){\line(1,0){150}}
\put(30,30){\mbox{$\ldots$}}

\put(50,50){\line(1,0){100}}
\put(60,60){\line(1,0){90}}
\put(70,70){\mbox{$\ldots$}}

\put(0,0){\line(0,1){140}}
\put(10,10){\line(0,1){130}}
\put(20,20){\line(0,1){90}}

\put(50,50){\line(0,1){40}}
\put(60,60){\line(0,1){30}}

\put(0,140){\line(1,0){10}}
\put(0,130){\line(1,0){10}}
\put(0,120){\line(1,0){10}}
\put(0,110){\line(1,0){20}}
\put(0,100){\line(1,0){20}}
\put(0,90){\line(1,0){20}}
\put(0,80){\line(1,0){20}}
\put(0,70){\line(1,0){20}}
\put(0,60){\line(1,0){20}}
\put(0,50){\line(1,0){20}}
\put(0,40){\line(1,0){20}}
\put(0,30){\line(1,0){20}}
\put(0,20){\line(1,0){20}}
\put(0,10){\line(1,0){10}}

\put(50,90){\line(1,0){10}}
\put(50,80){\line(1,0){10}}
\put(50,70){\line(1,0){10}}
\put(50,60){\line(1,0){10}}

\put(200,0){\line(0,1){10}}
\put(190,0){\line(0,1){10}}
\put(180,0){\line(0,1){10}}
\put(170,0){\line(0,1){20}}
\put(160,0){\line(0,1){20}}
\put(150,0){\line(0,1){20}}
\put(140,0){\line(0,1){20}}
\put(130,0){\line(0,1){20}}
\put(120,0){\line(0,1){20}}
\put(110,0){\line(0,1){20}}
\put(100,0){\line(0,1){20}}
\put(90,0){\line(0,1){20}}
\put(80,0){\line(0,1){20}}
\put(70,0){\line(0,1){20}}
\put(60,0){\line(0,1){20}}
\put(50,0){\line(0,1){20}}
\put(40,0){\line(0,1){20}}
\put(30,0){\line(0,1){20}}
\put(20,0){\line(0,1){20}}
\put(10,0){\line(0,1){10}}

\put(60,50){\line(0,1){10}}
\put(70,50){\line(0,1){10}}
\put(80,50){\line(0,1){10}}
\put(90,50){\line(0,1){10}}
\put(100,50){\line(0,1){10}}
\put(110,50){\line(0,1){10}}
\put(120,50){\line(0,1){10}}
\put(130,50){\line(0,1){10}}
\put(140,50){\line(0,1){10}}
\put(150,50){\line(0,1){10}}

\qbezier(-5,5)(-25,70)(-5,135)
\put(-30,67){\mbox{$a_1$}}
\qbezier(5,-5)(100,-35)(195,-5)
\put(95,-30){\mbox{$b_1$}}

\qbezier(45,55)(35,70)(45,85)
\put(25,67){\mbox{$a_k$}}
\qbezier(55,45)(100,25)(145,45)
\put(95,25){\mbox{$b_k$}}

\put(2,2){\mbox{$1$}}
\put(12,12){\mbox{$2$}}
\put(52,52){\mbox{$k$}}

\end{picture}

\noindent
Remarkably, (\ref{Ss}) can be rewritten in these coordinates as another determinant representation,
where the entries are the single-hook characters\footnote{This formula reflects a peculiarity of the Schur functions: $\chi_{(\vec\alpha|\vec\beta)}$: depends on the first $\alpha_1+\beta_1-1$ time variables $p_k$. Deformations of the Schur functions, for instance, to the Macdonald polynomials do not celebrate this property at any finite values of $t$ and $q$ but $t=q$: the polynomials become dependent on all first $\sum_i^n(\alpha_i+\beta_i)-n$ time variables.}, \cite{KM,DJKMIII}:
\be\label{Sh}
\chi_{(\vec\alpha|\vec\beta)}=\det_{i,j}\chi_{(\alpha_i|\beta_j)}
\ee
According to (\ref{Ss}),
the one-hook character is expressed through the symmetric characters as
\be
\chi_{(a|b)}=(-1)^{b}\sum_{j\ge 0}\chi_{[j+b]}\{-p_k\}\chi_{[a-j-1]}\{p_k\}=
(-1)^{b+1}\sum_{j\ge 0}\chi_{[b-j-1]}\{-p_k\}\chi_{[a+j]}\{p_k\}
\ee
thus (\ref{Sh}) is a somewhat ``weaker" version of (\ref{Ss}).
Instead it has nicer properties and a broader variety of applications
and deformations.
In fact, there is a formula that mixes two representations (\ref{Sh}) and (\ref{Ss}),
but at this moment it looks heavier, see \cite[eq.(24)]{DJKMIII},
and lacks clear applications.

\section{Skew characters in the hook variables}

For the skew Schur functions $\chi_{R/P}$ with $R$ being one-hook, there is a very simple formula in the Frobenius coordinates:
\be\label{sk1}
\chi_{(a|b)/(c|d)}=\chi_{(a-c|1)}\chi_{(1|b-d)}
\ee
Since all Young diagrams with more than one hook are not contained in the one-hook diagram, all other skew Schur functions vanish in this case. In fact, this is a corollary of the corresponding property of any skew Schur function: if the Young diagram $P$ being embedded into the diagram $R$ in such a way that their origins coincide, parts $R$ into two untied Young diagrams $R_1$ and $R_2$, then
\be
\chi_{R/P}=\chi_{R_1}\chi_{R_2}
\ee
Formula (\ref{sk1}), in particular, implies that
\be\label{hchi}
\hat\chi_{(a|b)}\cdot\chi_{(c|d)}=\chi_{(a-c|1)}\chi_{(1|b-d)}
\ee
where $\hat\chi_R:=\chi_R\{{1\over k}{\partial\over \partial p_k}\}$. Let us introduce a new operator $\hat D_{(c|d)}$ acting on the single-hook Schur functions in such a way that
\be
\hat D_{(c|d)}\cdot\chi_{(a|b)}:= \chi_{(a-c|1)}\chi_{(1|b-d)}
\ee
Now we can write down what is the skew Schur function when $P$ is single-hook, $P=(c|d)$. In this case,
\be\label{10}
\chi_{(\vec\alpha|\vec\beta)/(c|d)}=\hat D_{(c|d)}\cdot\det_{i,j}\chi_{(\alpha_i|\beta_j)}
\ee
where the operator $\hat D_{(c|d)}$ acts on the products of the single-hook Schur functions in the determinant by the Leibnitz rule, which is non-trivial, since the operator $\hat\chi_{(a|b)}$ in (\ref{hchi}) contains higher derivatives. In other words,
\be
\chi_{(\vec\alpha|\vec\beta)/(c|d)}=\chi_{(\vec\alpha|\vec\beta)}\cdot\sum_{i,j}\Big(\chi^{-1}\Big)_{\alpha_i,\beta_j}
\chi_{(\alpha_j-c|1)}\chi_{(1|\beta_i-d)}
\ee
where $\chi^{-1}$ is the matrix inverse to $\chi_{(\alpha_i|\beta_j)}$ and we used Jacobi's formula for the derivative of the determinant \cite{Jacobi}
\be
\hat D\cdot\det \chi=\det \chi\cdot\Tr\Big(\chi^{-1}\cdot\hat D\chi\Big)
\ee
Using this formula, one can similarly describe the action of a few successive operators $\hat D$ in a similar way (note that all of them commute). For instance, for $\hat D_1=\hat D_{(a|b)}$, $\hat D_2=\hat D_{(c|d)}$ with $a,b,c,d>1$,
\be
\hat D_2\hat D_1\cdot\det\chi=\det\chi\cdot\Tr \Big(\chi^{-1}\cdot\hat D_1\chi\Big)\cdot\Tr \Big(\chi^{-1}\cdot\hat D_2\chi\Big)-
\det\chi\cdot\Tr\Big(\chi^{-1}\cdot\hat D_1\chi\cdot\chi^{-1}\cdot\hat D_2\chi\Big)
\ee
etc.

One can check that the action of a product of operators $\hat D$ celebrates a simple property
\be
\ldots\hat D_{(a_1,b_1)}\hat D_{(a_2,b_2)}\ldots\cdot\chi=-\ldots\hat D_{(a_1,b_2)}\hat D_{(a_2,b_1)}\ldots\cdot\chi
\ee
i.e. it depends only on the two sets $\{a_i\}$ and $\{b_i\}$ and the parity of the permutation. Hence,
\be
\Big(\prod_i^n\hat D_{(a_i,b_i)}\Big)\cdot\chi={1\over n!}\det_{i,j}\hat D_{(a_i|b_j)}\cdot\chi
\ee
Now everything is ready for our final formula for the generic skew Schur function:
\be\boxed{
\chi_{(\vec\alpha|\vec\beta)/(\vec\gamma|\vec\delta)}=\Big(\prod_i^n\hat D_{(\gamma_i,\delta_i)}\Big)\cdot
\det_{i,j}\chi_{(\alpha_i|\beta_j)}
}
\ee
The r.h.s. of this formula can be also rewritten in the matrix form:
\be
\Big(\prod_i^n\hat D_{(\gamma_i,\delta_i)}\Big)\cdot\det\chi=\det\chi\cdot\sum_{a_1,\ldots,a_n}\sum_{\sigma\in S_n}(-1)^{p_\sigma}
Z^{(\gamma_1|\delta_1)}_{\alpha_{a_1},\beta_{a_{\sigma(1)}}}\ldots
Z^{(\gamma_n|\delta_n)}_{\alpha_{a_n},\beta_{a_{\sigma(n)}}}
\ee
where the matrix $Z_{\alpha_a,\beta_b}^{(\gamma_i|\delta_i)}:=\Big(\chi^{-1}\cdot\hat D_{(\gamma_i|\delta_i)}\chi\Big)_{a,b}
%=\sum_c\chi^{-1}_{\alpha_a,\beta_c}\cdot\chi_{(\alpha_c-\gamma_i|1)}\cdot\chi_{(1|\beta_b-\delta_i)}
$ and the sum goes over all repeated indices and over all permutations from the symmetric group $S_n$, $p_\sigma$ being the parity of permutation $\sigma$.

\section{$\tau$-functions and Schur functions\label{stau}}

It is well-known that the linear and bilinear combinations of the Schur functions may give rise to a $\tau$-function: in order for a linear combination
\be\label{lc}
\sum_R w_R\chi_R\{p\}
\ee
to be a $\tau$-function of the KP hierarchy\footnote{Note that usually in integrable theory other time variables are used: $t_k=p_k/k$.}, the coefficients in front of the Schur functions should satisfy the Pl\"ucker relations (see (\ref{Pr}) below), \cite{DJKM,Sat}. Similarly, there are explicit conditions for $w_{RR'}$ \cite{Taka} that a bilinear combination
\be\label{bc}
\sum_{R,R'}w_{RR'}\chi_R\{p_k\}\chi_{R'}\{p'_k\}
\ee
is a $\tau$-function of the Toda hierarchy, or a KP $\tau$-function w.r.t. the both sets of time variables $\{p_k\}$ and $\{p'_k\}$. This latter $\tau$-function can be also treated as a $\tau$-function of the 2-component KP hierarchy. However, no extension to multilinear combinations of the Schur functions that is still a $\tau$-function has been known so far.

Note that there is a relatively simple class of coefficients $w_{R}$ and $w_{RR'}$ that gives rise to the $\tau$-functions \cite{GKM2,Ok,OS,AMMN1,AMMN2}:
\be
w_{RR'}=\delta_{RR'}\prod_{i,j\in R}f(i-j),\ \ \ \ \ \ \ \ \ \ w_{R}=\prod_{i,j\in R}f(i-j)
\ee
where $f(x)$ is an arbitrary function. Such $\tau$-functions are called hypergeometric \cite{OS}. In the case of $f(x)=1$, the sum (\ref{bc}) is immediately calculated using the Cauchy formula
\be
\sum_{R}\chi_R\{p_k\}\chi_{R}\{p'_k\}=\exp\left(\sum_k {p_kp'_k\over k}\right)
\ee
and is a trivial $\tau$-function.

In \cite{IMM,IMMint}, we suggested a possible extension of this Cauchy formula to the multilinear case
\be
\sum_{R_1,\ldots,R_r}\mathfrak{C}_{_{R_1\ldots R_r}} \prod_{m=1}^r\chi_{R_m}\{p^{(m)}\}
=\exp\left(\sum_k
{\prod_{m=1}^rp^{(m)}_k\over k}\right),\nn\\
\mathfrak{C}_{R_1,\ldots,R_r}:=\sum_{\Delta\vdash n}{\prod_{i=1}^r \psi_{R_i}(\Delta)\over z_\Delta}
\ee
which gives rise to a (still trivial) KP $\tau$-function w.r.t. all time variables and can be further made slightly less trivial \cite{IMMint}.

A the same time, A. Orlov proposed another multilinear extension of (\ref{bc}), which we discuss in the next section.

\section{Hook variables and Matisse $\tau$-functions\label{Matisse}}

 Using the Frobenius coordinates, consider the sum
\be\label{1}
\tau_0^{(r)}:=\sum_{\vec\alpha_i} \prod_{i=1}^r\chi_{(\vec\alpha_i|\vec\alpha_{i+1})}\{p^{(i)}\}
\ee
It is a KP $\tau$-function w.r.t. any of the time variables if one imposes the condition $\vec\alpha_{r+1}=\vec\alpha_{1}$. Hereafter, we imply the ordering $\alpha_1>\alpha_2>\ldots>0$ of the vector components, and the sum over $\vec\alpha$ means a sum over all possible ordered values of the vector components, and over all possible numbers $l$ of non-zero components, i.e. over all possible sets satisfying the ordering condition.

Moreover, the expression
\be\label{2}
\tau^{(r)}:=\sum_{\vec\alpha_i} \prod_{k=1}^r\left(\chi_{(\vec\alpha_i|\vec\alpha_{k+1})}\{p^{(k)}\}\prod_{i,j\in (\vec\alpha_k|\vec\alpha_{k+1})}
f_k(i-j)\right)
\ee
is still a $\tau$-function, $f_k$ being arbitrary functions\footnote{This is because any function $w_R$ that satisfies the Pl\"ucker relations can be multiplied $w_R\longrightarrow w_R\prod_{i,j\in R}f(i-j)$ by an arbitrary function $f(x)$ and still continues to satisfy the Pl\"ucker relations \cite{GKM2,OS}: these latter are invariant with respect to this operation.}. Such $\tau$-functions have been called by A.Orlov ``Matisse $\tau$-functions" (because of the ``Dance" painting \cite{Mat}).

In order to prove that (\ref{1}) is a $\tau$-function w.r.t. the variables $p_k$, it is sufficient to prove that all the Pl\"ucker relations
\be\label{Pr}
w_{(\vec\alpha|\vec\beta)[\alpha_i,\alpha_j;\beta_i,\beta_j]}\cdot w_{(\vec\alpha|\vec\beta)}-
w_{(\vec\alpha|\vec\beta)[\alpha_i;\beta_i]}\cdot w_{(\vec\alpha|\vec\beta)[\alpha_j;\beta_j]}+
w_{(\vec\alpha|\vec\beta)[\alpha_i;\beta_j]}\cdot w_{(\vec\alpha|\vec\beta)[\alpha_j;\beta_i]}=0
\ee
for the coefficients $w_{(\vec\alpha|\vec\beta)}$ of $\chi_{(\vec\alpha|\vec\beta)}\{p_k\}$ in (\ref{1}) are satisfied. Here we denoted through $[\{\alpha_i\};\{\beta_j\}]$ removing a subset $\{\alpha_i\};\{\beta_j\}$ from the set of hook legs and arms.

The simplest way to prove this is as follows.
We can use (\ref{Sh}) and apply the formula
\be\label{3}
\det_{i,j}\sum_{k=1}^\infty A_{ik}B_{kj}=\sum_{1\le k_1<k_2<\ldots}\det_{ij}A_{ik_j}\det_{ij}B_{k_ji}
\ee
Consider, for instance, $r=3$. Then, (\ref{1}) is a KP $\tau$-function if
\be
w_{(\vec\alpha|\vec\beta)}=\sum_{\vec\gamma}\chi_{(\vec\beta|\vec\gamma)}\chi_{(\vec\gamma|\vec\alpha)}
\ee
satisfies the Pl\"ucker relations. Using (\ref{3}), this formula is reduced to
\be\label{4}
w_{(\vec\alpha|\vec\beta)}=\det_{ij}\left(\sum_{k=1}^\infty\chi_{(\beta_i|k)}\chi_{(k|\alpha_j)}\right)
\ee
again with one-hook determinant entries, and one can trivially check that it satisfies the Pl\"ucker relations: they are just the Jacobi identity\footnote{If we denote through
$\displaystyle{J\left(\begin{array}{lcr}
i_1&\ldots&i_s\cr
j_1&\ldots&j_s
\end{array}\right)}$
the determinant of a matrix with removed rows $i_1,\ldots,i_s$ and columns $j_1,\dots,j_s$, then the Jacobi identity reads
\be
J\left(\begin{array}{lr}
i& j\cr
i& j
\end{array}\right)\cdot J=J\left(\begin{array}{c}
i\cr
i
\end{array}\right)\cdot J\left(\begin{array}{c}
j\cr
j
\end{array}\right)-J\left(\begin{array}{c}
i\cr
j
\end{array}\right)\cdot J\left(\begin{array}{c}
j\cr
i
\end{array}\right)
\ee
 } for the determinant (\ref{4}).

Because of (\ref{3}), it is natural to introduce the kernel
\be
{\cal K}^{(r)}_{ij}\{p^{(m)}\}:=\chi_{(i|k_2)}\{p^{(1)}\}\left(\prod_{m=2}^{r-1}\chi_{(k_m|k_{m+1})}\{p^{(m)}\}\right)
\chi_{(k_r|j)}\{p^{(r)}\}
\ee
Then, using (\ref{3}), one can present (\ref{1}) in the form
\be\label{tau}
\tau_0^{(r)}:=\sum_{\vec\alpha}\det_{ij}{\cal K}^{(r)}_{\alpha_i,\alpha_j}\{p^{(m)}\}
\ee
Moreover, in order to prove that (\ref{1}) is a KP $\tau$-function, one has to check that
\be
w_{(\vec\alpha|\vec\beta)}=\det_{ij}{\cal K}^{(r-1)}_{\alpha_i,\beta_j}\{p^{(m)}\}
\ee
satisfies the Pl\"ucker relations, which is, indeed, the case because of the Jacobi identity.

Note that, for any matrix $A_{ij}$, the following formula is correct:
\be
\sum_{\vec\alpha}\det_{ij}A_{\alpha_i,\alpha_j}=\sum_{n=0}^\infty\chi_{_{[1^n]}}\{p_k=\Tr A^k\}=
\exp\left(\sum_n{(-1)^{n+1}\over n}\Tr A^n\right)
\ee
since $\chi_R\{p_k\}=\chi_{_{R^\vee}}\{(-1)^{k+1}p_k\}$ and $R^\vee$ denotes the transposed Young diagram.
It follows from the formula
\be
\sum_{\alpha_1>\ldots>\alpha_n}\det_{ij}A_{\alpha_i,\alpha_j}=\chi_{_{[1^n]}}\{p_k=\Tr A^k\}
\ee
Then, denoting through $K$ the matrix with matrix elements ${\cal K}^{(r)}_{ij}$, one can rewrite (\ref{tau}) as
\be
\tau_0^{(r)}=\exp\left(\sum_n{(-1)^{n+1}\over n}\Tr K^n\right)
\ee
One can also consider the matrix $X$ with matrix elements $X_{ik}^{(m)}:=\chi_{(i|k)}\{p^{(m)}\}$ and write equivalently
\be\boxed{
\tau_0^{(r)}=\exp\left(\sum_n{(-1)^{n+1}\over n}\Tr \Big(X^{(1)}X^{(2)}\ldots X^{(r)}\Big)^n\right)}
\ee
This formula remains correct for (\ref{2}) with $X_{ik}^{(m)}=\chi_{(i|k)}\{p^{(m)}\}\cdot \prod_{i,j\in (i|k)}f_m(i-j)$.

Let us consider the simplest case of $r=2$. Then,
\be
{\cal K}_{ij}=\sum_k\chi_{(i|k)}\{p^{(1)}\}
\chi_{(k|j)}\{p^{(2)}\}
\ee
and\footnote{If one defines
\be
{\cal K}_{ij}'=\sum_k\chi_{(i|k)}\{p^{(1)}\}
\chi_{(j|k)}\{p^{(2)}\}
\ee
this formula gets the form
\be
\sum_{n=1}^\infty{(-1)^n\over n}\Tr K'^n=\sum_{n=1}^\infty{p^{(1)}_n p^{(2)}_n\over n}
\ee
}
\be\boxed{
\sum_{n=1}^\infty{(-1)^n\over n}\Tr K^n=\sum_{n=1}^\infty (-1)^{n}{p^{(1)}_n p^{(2)}_n\over n}}
\ee
This formula is a corollary of the fact that, as immediately follows from (\ref{1}),
\be
\tau_0^{(2)}=\exp\left(\sum_{n=1}^\infty (-1)^{n+1}{p^{(1)}_n p^{(2)}_n\over n}\right)
\ee

\section{Hook variables and shifted Schur functions}

The shifted Schur functions are unambiguously defined by the formula
\be\label{sSe}
\chi^*_\mu\{p_k\}=\chi_\mu\{p\}+\sum_{\lambda:\ |\lambda|<|\mu|} c_{\mu\lambda}\chi_\lambda\{p_k\}
\ee
in such a way that the coefficients are (unambiguously) determined from the conditions
\be
\chi^*_\mu\{^*p_k(R)\}=0\ \ \ \ \ \ \ \ \hbox{if }\mu\notin R\nn\\
\nn\\
^*p_k(R):=\sum_i \left[(R_i-i)^k-(-i)^k\right]
\ee
They are symmetric functions of the variables $x_i-i$, $i=1,\ldots,N$ on the locus formed by the shifted power sums
\be
^*p^{(N)}_k(x):=\sum_{i=1}^N \left[(x_i-i)^k-(-i)^k\right]
\ee

These shifted Schur functions have three interesting properties:
\begin{itemize}
\item
Surprisingly, these functions have a similar determinant representation, the determinant entries being one-hook characters:
\be
\chi^*_{(\vec\alpha|\vec\beta)}=\det_{i,j}\chi^*_{(\alpha_i|\beta_j)}
\ee
At the same time, formula (\ref{Ss}) is not immediately generalized to the case of shifted functions.
\item
The shifted one-hook Schur function $\chi^*_R$ is given by a linear combination of only one-hook Schur functions $\chi_Q$ such that $Q\in R$:
\be
\chi^*_{(\alpha|\beta)}\{p_k\}=\sum_{i=1}^{\alpha}\sum_{j=1}^{\beta}C_{\alpha-1,i-1}\cdot C_{\beta,j}\cdot
\chi_{(i|j)}\{p_k\}
\ee
where the coefficients $C_{ij}$ are defined by the expansion of the Pochhammer symbol
\be
\prod_{i=0}^{k-1}(z-i)=\sum_{j=1}^k C_{kj}z^j
\ee
\item
{\bf The shifted Schur functions themselves satisfy the Hirota bilinear equations in $p^*_k$}, i.e. the coefficients $c_{\mu\lambda}$ in (\ref{sSe}) satisfy the Pl\"ucker relations in $\lambda$ at any $\mu$. However, even if the coefficients of a linear combination of the {\it shifted} Schur functions satisfy the Pl\"ucker relations, this does not give rise to a KP $\tau$-function.
\end{itemize}
Hence, all the formulas of the previous section still persist, but the counterpart of (\ref{1})-(\ref{2}) is no longer a $\tau$-function.

\section{Cut-and-join operators in the hook variables}

First of all, let us note that the time variables $p_a$ are expanded only into single-hook Schur functions
\be\label{p}
p_a=\sum_{i=1}^{a}(-1)^{i+1}\chi_{(a-i+1|i)}
\ee
Hence, the products of time variables $p_ap_b$ are expanded into double-hook Schur functions,  $p_ap_bp_c$, into triple-hook Schur functions, etc. For instance, for $a\ge b$
\be\label{pp}
p_ap_b=(-1)^{a+b}\sum_{j=1}^b\sum_{i=1}^{a-b-1}(-1)^i\chi_{(i+j,j|a+1-i-j,b+1-j)}+\nn\\
+\sum_{{i\ge 1,j\ge 1\atop i+j\le b}}(-1)^{i+j+b}\chi_{(a+1-j,i|b+1-i,j)}
+\sum_{{i\ge 1,j\ge 1\atop i+j\le b}}(-1)^{i+j+a}\chi_{(b+1-j,i|a+1-i,j)}+\nn\\
+\sum_{i=1}^b(-1)^{i+1}\Big(\chi_{(a+b+1-i|i)}+(-1)^{a+b}\chi_{(i|a+b+1-i)}\Big)
\ee
These formulas can be used for constructing the cut-and-join operators \cite{MMN}
\be
\hat W_{\!_\Delta} = \frac{1}{z_{_\Delta}}:\prod_i \hat D_{\delta_i}:
\label{Wops}
\ee
where
\be
\hat D_k = \Tr (M \p_{M})^k
\ee
and $M$ is a matrix. The normal ordering in (\ref{Wops}) implies that all the derivatives $\p_M$
stand to the right of all $M$. We apply $W_\Delta$ only to
gauge invariants, and they are themselves ``gauge"-invariant matrix operators, thus, they can be realized as differential operators in $p_k = \Tr M^k$. The Schur functions form a system of common eigenfunctions of the cut-and-join operators \cite{MMN},
\be
\hat W_{\!_\Delta} \chi_R\{p\} = \phi_R(\Delta)\cdot \chi_R\{p\}
\label{evW}
\ee
with the eigenvalues \cite{IK}
\be
\phi_R(\Delta)=\sum_{\mu\vdash |\Delta|} \chi^*_\mu\Big({^*p_k}(R)\Big)\ {\psi_\mu(\Delta)\over z_\Delta}
\ee
First few examples of the cut-and-join operators in terms of the time-variables are
\be
\hat{  W}_{[1]} = \tr \hat D= \sum_{k=1} kp_k\frac{\p}{\p p_k}
\ee
\be
\hat{  W}_{[2]} ={1\over 2} \, :\tr \hat D^2\,:\ =
\frac{1}{2}\sum_{a,b=1}^\infty
\left( (a+b)p_ap_b\frac{\p}{\p p_{a+b}} +
abp_{a+b}\frac{\p^2}{\p p_a\p p_b}\right)
\ee
\be
\hat{  W}_{[3]} = \frac{1}{3}\, :\tr \hat D^3\,:\ =
\frac{1}{3}\sum_{a,b,c\geq 1}^\infty
abcp_{a+b+c} \frac{\p^3}{\p p_a\p p_b\p p_c}
+ \frac{1}{2}\sum_{a+b=c+d} cd\left(1-\delta_{ac}\delta_{bd}\right)
p_ap_b\frac{\p^2}{\p p_c\p p_d} + \\
+ \frac{1}{3} \sum_{a,b,c\geq 1}
(a+b+c)\left(p_ap_bp_c + p_{a+b+c}\right)\frac{\p}{\p p_{a+b+c}}
\label{W3p}
\ee
Since \cite{MMN}
\be
\hat{  W}_{[1,1]}={1\over 2}\hat{  W}_{[1]}(\hat{  W}_{[1]}-1)\\
\hat{  W}_{[2,1]}=\hat{  W}_{[2]}(\hat{  W}_{[1]}-2)\\
\hat{  W}_{[1,1,1]}={1\over 6}\hat{  W}_{[1]}(\hat{  W}_{[1]}-2)(\hat{  W}_{[1]}-1)
\ee
we are interested only in $\hat{  W}_{R} $ with $R$ that does not contain the unit cycle, i.e. in $\hat{  W}_{[1]} $, $\hat{  W}_{[2]} $, $\hat{  W}_{[3]} $, etc.

Now one can realize the cut-and-join operators in terms of hook variables. Let us introduce the Schur functions of derivatives
\be
\hat\chi_R:=\chi_R\Big\{k{\p\over\p p_k}\Big\}
\ee
Then, the cut-and-join operator can be realized as a sum
\be\label{Wd}
\hat W_{\!_\Delta}=\sum C^{(\Delta)}_{RP}\chi_R\hat\chi_P
\ee
with some coefficients $C^{(\Delta)}_{RP}$. For instance, the first non-trivial operators are realized as \cite{MMrev}
\be\label{W12}
\hat{  W}_{[1]} = \sum_{a=1} ap_a\frac{\p}{\p p_a}=\sum_{a=1}\sum_{s,s'=1}^{a}(-1)^{s+s'}\cdot\chi_{(a-s+1|s)}\cdot
\hat\chi_{(a-s'+1|s')}=
 \sum_{\stackrel{r,s,r',s'=1}{r+s=r'+s' }}^\infty
(-)^{s+s'} \cdot \chi_{(r|s)}\cdot\hat\chi_{(r'|s')}
\\
\hat{  W}_{[2]} =\frac{1}{2}\sum_{a,b=1}^\infty \Big((a+b)p_ap_b\p_{a+b} + ab\,p_{a+b}\p_a\p_b\Big)
= {1\over 2}\sum_{\stackrel{r,s,r',s'=1}{r+s=r'+s' }}^\infty
(-)^{s+s'} (r-s+r'-s') \cdot \chi_{(r|s)}\cdot\hat\chi_{(r'|s')}
\ee
where we used formulas (\ref{p}) and (\ref{pp}). Naively, one could expect basing on (\ref{pp}) that the $\hat{  W}_{[2]}$-operator depends on  2-hook Young diagrams. However, there is conspiracy, and only single-hook diagrams enter the result. Similarly, the $\hat{  W}_{\Delta}$-operators with higher $|\Delta|$ such that $\Delta$ does not contain cycles of unit length depend on Young diagrams with not more than $|\Delta|-1$ hooks (the number of hooks can be less: for instance, $\hat W_{[4]}$ involves not more than double-hook Young diagrams, $\hat W_{[5]}$ triple hooks, etc; see (\ref{cc})). This removing cycles of unit length is much similar to removing $U(1)$-factors from $U(N)$ algebras, which results in simple algebras of rank one unit less. For instance, the $\hat{  W}_{[3]}$-operator has the following hook form representation
\be
\hat{  W}_{[3]}={1\over 2} \sum_{\stackrel{r,s,r',s'=1}{r+s=r'+s' }}^\infty
(-)^{s+s'} C^{([3])}_{(r,s),(r',s')} \cdot \chi_{(r|s)}\cdot\hat\chi_{(r'|s')}+\nn\\
+\sum_{r_1+r_2+s_1+s_2=r'+s'+1}
C^{([3])}_{([r_1,r_2]|[s_1,s_2]),(r',s')}\cdot \left(\chi_{([r_1,r_2]|[s_1,s_2])}\cdot\hat\chi_{(r'|s')}+
\chi_{(r'|s')}\cdot\hat\chi_{([r_1,r_2]|[s_1,s_2])}\right)+\nn\\
+\sum_{\sum_{i=1}^2(r_i+s_i-r_i'-s'_i)=0}
C^{([3])}_{([r_1,r_2]|[s_1,s_2]),[r_1',r_2']|[s_1',s_2'])}\cdot \chi_{([r_1,r_2]|[s_1,s_2])}\cdot\hat\chi_{([r_1',r_2']|[s_1',s_2'])}
\ee
where
\be\label{60}
C^{([3])}_{(r,s),(r',s')}={1\over 2}\Big(r-s+r'-s'\Big)^2+(s'-r')\Theta(r-s-(r'-s'))+\nn\\
+(s-r)\Theta(r'-s'-(r-s)-2)
+\Theta(s+s'-r-r'-2)(r+r'-s'-s)-(r+s-2)
\ee
and $\Theta(x)$ is the Heaviside function, and the other coefficients $C^{([3])}_{([r_1,r_2]|[s_1,s_2]),(r',s')}$ and $C^{([3])}_{([r_1,r_2]|[s_1,s_2]),[r_1',r_2']|[s_1',s_2'])}$ are also rather involved.

\paragraph{A comment on evaluating the coefficients $C^{(\Delta)}_{RP}$.}
Note that, since
\be
\hat\chi_R\chi_P=\chi_{P/R}
\ee
one obtains from (\ref{evW}) that
\be\label{matrix}
\hat W_{\!_\Delta} \chi_R =\sum C^{(\Delta)}_{QP}\chi_Q\chi_{R/P}= \phi_R(\Delta)\cdot \chi_R
\ee
i.e. one can realize the $\hat W_{\!_\Delta}$-operator not as a differential operator, but as an operator acting on the linear space of Young diagrams. To this end, one can note that
\be
\chi_{R/P}=\sum_{S} N^R_{PQ}\chi_S
\ee
hence, the $\hat W_{\!_\Delta}$-operator can be given by an infinite matrix in the space of Young diagrams:
\be
\Big(\hat W_{\!_\Delta}\Big)_{RS}=\sum_{X,Y,Z}C^{(\Delta)}_{XY}N^S_{XZ}N^R_{YZ}
\ee
where $N^R_{XY}$ are the Littlewood-Richardson coefficients, and $|R|=|S|$. Here we used that
\be
\chi_X\{p\}\chi_Y\{p\}=\sum_ZN^Z_{XY}\chi_Z\{p\}\nn\\
\chi_{X/Y}=\sum_ZN^X_{YZ}\chi_Z
\ee
Thus, the eigenvalue equation (\ref{matrix}) can be rewritten in the matrix form as
\be
\sum_{{X,Y,Z\atop |X|=|Y|}}C^{(\Delta)}_{XY}N^S_{XZ}N^R_{YZ}=\phi_R(\Delta)\delta_{RS}
\ee
This linear system unambiguously defines a symmetric matrix $C^{(\Delta)}$ for any given function $\phi_R(\Delta)$.

In fact, the system of equations
\be\label{ls}
\sum_{{X,Y,Z\atop |X|=|Y|}}C_{XY}N^S_{XZ}N^R_{YZ}=\lambda_R\delta_{RS}
\ee
can be simplified if one makes the linear transform with the kernel $\displaystyle{\psi_X(\Delta)\psi_Y(\Delta')\over z_\Delta z_\Delta'}$:
\be
A_{XY}\longrightarrow \tilde A_{\Delta\Delta'}:=\sum_{{X,Y\atop |X|=|Y|}}{\psi_X(\Delta)\psi_Y(\Delta')\over z_\Delta z_\Delta'}A_{XY}
\ee
Then, using
\be
N^Z_{XY}=\sum_{\Delta_1,\Delta_2}{\psi_Z(\Delta_1+\Delta_2)\psi_X(\Delta_1)\psi_Y(\Delta_2)\over z_{\Delta_1}z_{\Delta_2}}
\ee
where the sum of two Young diagrams is understood as a reordered unification of rows from the both diagrams, one can rewrite (\ref{ls}) in the form
\be
\sum_{{\Delta_1,\Delta_2\atop \Delta-\Delta_1=\Delta'-\Delta_2}}{\tilde C_{\Delta_1\Delta_2}\over z_{\Delta-\Delta_1}}=
\tilde\lambda_{\Delta\Delta'}
\ee

\section{Single-hook $\hat W$-operators\label{s7}}

Let us now understand how many $\hat W^{sh}$-operators exist that involve only single-hook diagrams in expansion (\ref{Wd}). The eigenfunction condition (\ref{evW}) does not restrict $\hat W^{sh}$ when it is acting on the single-hook diagrams. Consider its action on the double-hook diagrams. Then, as it follows from (\ref{10}),
\be
\hat W^{sh}\Big(\chi_{(r_1,s_1)}\chi_{(r_2,s_2)}-\chi_{(r_1,s_2)}\chi_{(r_2,s_1)}\Big)=\nn\\
=\lambda_{(r_1,s_1)}\chi_{(r_1,s_1)}\chi_{(r_2,s_2)}+
\lambda_{(r_2,s_2)}\chi_{(r_1,s_1)}\chi_{(r_2,s_2)}-
\lambda_{(r_1,s_2)}\chi_{(r_1,s_2)}\chi_{(r_2,s_1)}-\lambda_{(r_2,s_1)}\chi_{(r_1,s_2)}\chi_{(r_2,s_1)}
\ee
i.e.
\be
\lambda_{([r_1,r_2],[s_1,s_2])}=\lambda_{(r_1,s_1)}+\lambda_{(r_2,s_2)}=\lambda_{(r_1,s_2)}+\lambda_{(r_2,s_1)}
\ee
At level 4 ($r_1=s_1=2$, $r_2=s_2=1$), it imposes two restrictions on the eigenvalues (and, hence, in accordance with (\ref{ls}), on the operator $\hat W^{sh}$):
\be
\lambda_{([2,1],[2,1])}=\lambda_{(2,2)}+\lambda_{(1,1)}=\lambda_{(2,1)}+\lambda_{(1,2)}
\ee
Similarly, at level 5, there are four more restrictions:
\be\label{c2}
\lambda_{([3,1],[2,1])}=\lambda_{(3,2)}+\lambda_{(1,1)}=\lambda_{(3,1)}+\lambda_{(1,2)}\nn\\
\lambda_{([2,1],[3,1])}=\lambda_{(2,3)}+\lambda_{(1,1)}=\lambda_{(2,1)}+\lambda_{(1,3)}
\ee
at level 6, eight additional restrictions, and at level $k$, $p(k)-3$ additional restrictions, $p(k)$ being the number of partitions of $k$.

It looks like still there is a large ambiguity in allowed eigenvalues $\lambda_R$. However, starting from level 9, there appear triple-hook diagrams, etc, which imposes more conditions. For instance, the triple-hooks diagrams impose seven restrictions for eigenvalues for every triple-hook diagram:
\be\label{c3}
\lambda_{([r_1,r_2,r_3],[s_1,s_2,s_3])}=\lambda_{(r_1,s_1)}+\lambda_{(r_2,s_2)}+\lambda_{(r_3,s_3)}=
\lambda_{(r_1,s_2)}+\lambda_{(r_2,s_3)}+\lambda_{(r_3,s_1)}=\lambda_{(r_1,s_3)}+\lambda_{(r_2,s_1)}+\lambda_{(r_3,s_2)}
=\nn\\
=\lambda_{(r_1,s_3)}+\lambda_{(r_2,s_2)}+\lambda_{(r_3,s_1)}
=\lambda_{(r_1,s_1)}+\lambda_{(r_2,s_3)}+\lambda_{(r_3,s_2)}=
\lambda_{(r_1,s_2)}+\lambda_{(r_2,s_1)}+\lambda_{(r_3,s_3)}
\ee
The conditions for double-hook diagrams remain only $3k-3$ independent eigenvalues up to level $k$, which gives 30 for the levels up to 11. At the same time, there are 8 triple-hook diagrams up to this level, which gives 56 additional conditions. Hence, normally, one would not expect any freedom in eigenvalues remaining.

It remains to notice that these conditions (\ref{c2}), (\ref{c3}), etc are solved by any eigenvalues of the form
\be\label{evc}
\boxed{
\lambda_{(\vec\alpha|\vec\beta)}=\sum_i \Big(\xi_1(\alpha_i)+\xi_2(\beta_i)\Big)
}
\ee
with arbitrary functions $\xi_{1,2}(x)$. Linear functions give rise to $\hat{  W}_{[1]}$ and quadratic, to $\hat{  W}_{[2]}$, since
\be
\phi_{(\vec\alpha|\vec\beta)}([1])=\sum_i\Big(\alpha_i+\beta_i-1\Big),\ \ \ \ \ \ \ \xi_{1,2}(x)=x-1/2\nn\\
\phi_{(\vec\alpha|\vec\beta)}([2])=\sum_i\Big(\alpha_i^2-\alpha_i-\beta_i^2+\beta_i\Big),\ \ \ \ \ \ \ \ \xi_1(x)=-\xi_2(x)=x^2-x
\ee
At the same time,
\be
\phi_{(\vec\alpha|\vec\beta)}([3])+{1\over 2}\phi_{(\vec\alpha|\vec\beta)}([1])^2=
\sum_i \Big({1\over 3}\alpha_i^3-{1\over 2}\alpha_i^2+{2\over 3}\alpha_i+
{1\over 3}\beta_i^3-{1\over 2}\beta_i^2+{2\over 3}\beta_i-{1\over 2}\Big),\nn\\
\xi_1(x)=\xi_2(x)={1\over 3}x^3-{1\over 2}x^2+{2\over 3}x-{1\over 4}
\ee
This means that the combination $\hat{  W}_{[3]}+{1\over 2}\hat{  W}_{[1]}^2$ is a single-hook operator too. Moreover, it has simple expansion coefficients $C$ (cf. with (60)):
\be
\hat W^{sh}_3=\hat{  W}_{[3]}+{1\over 2}\hat{  W}_{[1]}^2 = {1\over 4}\sum_{\stackrel{r,s,r',s'=1}{r+s=r'+s' }}^\infty
(-)^{s+s'}\Big( (r-s+r'-s')^2+2\Big) \cdot \chi_{(r|s)}\cdot\hat\chi_{(r'|s')}
\ee
Other combinations, with distinct (linear, cubic) functions $\xi_1(x)$ and $\xi_2(x)$, or coinciding quadratic functions for
$\hat W^{sh}$ can not be made of the Casimir operators. Since these latter form a basis in the space of differential operators of finite order (local operators), such $\hat W^{sh}$ are given by non-local operators.

Note that, at each level $n$, there is exactly one new operator $\hat W^{sh}$, it is equal to $\hat W_{[n]}+Pol_{n-1}(W_{\Delta})$, where $Pol_k$ denotes a polynomial, which is a sum of monomials $\prod_i\hat W_{\Delta_i}$ such that $\sum_i|\Delta_i|\le k$. In the next section, we construct a simple basis of such polynomials.

\section{Integrability and $\hat W$-operators}

As we discussed in sec.\ref{stau}, a diagonal bilinear combination of the Schur functions,
\be
\sum_Rf_R\chi_R\{p\}\chi_R\{p'\}
\ee
is a (hypergeometric) KP $\tau$-function w.r.t to both time variables $p_k$ and $\bar p_k$ iff $f_R=\prod_{i,j\in R}f(N-i+j)$  for arbitrary function $f(x)$ (it is also a Matisse $\tau$-function). Moreover, it is a Toda lattice $\tau$-function \cite{Taka,OS,AMMN2}. There is a more general expression, a generating function of the Hurwitz numbers \cite{Ok,MMN} (we called it Hurwitz $\tau$-function) with $f_R=\exp\left(\sum_{\Delta}\beta_\Delta\phi_R(\Delta)\right)$:
\be
Z^H=\sum_R\chi_R\{p\}\chi_R\{p'\}\exp\left(\sum_{\Delta}\beta_\Delta\phi_R(\Delta)\right)
\ee
where $\beta_\Delta$ are arbitrary constants labeled by Young diagrams. 
This generating function can be reproduced from the trivial exponential $\tau$-function $\exp\Big(\sum_{k=1}{p_k\bar p_k\over k}\Big)$
by action of the $\hat W$-operators:
\be
Z^H=\exp\left(\sum_\Delta\beta_\Delta\cdot\hat W_\Delta\right)\cdot\exp\Big(\sum_{k=1}{p_k\bar p_k\over k}\Big)
\ee
and it generally does not give rise to ordinary integrability. The point is that the basis of the commuting operators $\hat W_\Delta$ is too large: the cut-and-join operators form an additive basis of commuting operators, and, for ordinary integrability, one needs a multiplicative basis. What is this restricted basis?

A crucial property of the function $f_R=\prod_{i,j\in R}f(i-j)$ is that an exponential of an arbitrary linear combination of $C_n=\sum_i\left((R_i-i)^n-(-i)^n\right)$, which are eigenvalues of the $SL(N)$ Casimir operators, is of this kind \cite{AMMN2}. Hence,
\be\label{tauH}
\tau^H_N=\sum_R\chi_R\{p_k\}\chi_R\{\bar p_k\}\exp\left(\sum_n \zeta_nC_n(R)\right)
\ee
where $\zeta_n$ are arbitrary coefficients, is a KP $\tau$-function. We choose slightly different definition (linear combination) of $C_n$ that gives most simply looking formulas:
\be\label{cc}
\boxed{\begin{array}{rcl}
C_n(R):=\displaystyle{{1\over n}\sum_i\left[\Big(R_i-i+1/2\Big)^n-\Big(-i+1/2\Big)^n\right]}&=&\sum_i\Big(\xi_n(\alpha_i)+
(-1)^{n+1}\xi_n(\beta_i)\Big)
\\
&&
\\
\xi_n(x):&=&\displaystyle{(2x-1)^n\over 2^n\cdot n}
\end{array}
}
\ee
Every $C_n(R)$ is definitely a low triangle (in grading) polynomial combination of the eigenvalues $\phi_R(\Delta)$, the concrete combinations (\ref{cc}) being called completed cycles \cite{cc}. These combinations
can be realized either at the level of eigenvalues, or at the level of $\hat W_\Delta$-operators. For instance,
\be\label{cc}
\hat C_2=\hat{  W}_{[2]}\nn\\
\hat C_3=\hat{  W}_{[3]}+{1\over 2}\hat{  W}_{[1]}^2-{5\over 12}\hat{  W}_{[1]}\nn\\
\hat C_4=\hat{  W}_{[4]}+2\hat{  W}_{[2]}\hat{  W}_{[1]}-{11\over 4}\hat{  W}_{[2]}\nn\\
\hat C_5=\hat{W}_{[5]}+3\hat{W}_{[3]}\hat{W}_{[1]}+2\hat{W}_{[2]}^2-{19\over 2}\hat{W}_{[3]}
+{2\over 3}\hat{W}_{[1]}^3-{9\over 4}\hat{W}_{[1]}^2+{383\over 240}\hat{W}_{[1]}\nn\\
\ldots
\ee
They are linear combinations of the single-hook operators of the previous section and have a simple hook representation
\be\label{Cas}
\boxed{
\hat C_n = {1\over 2^n\cdot n}\sum_{\stackrel{r,s,r',s'=1}{r+s=r'+s' }}^\infty
(-)^{s+s'}\Big( (r-s+r'-s'+1)^n-(r-s+r'-s'-1)^n\Big) \cdot \chi_{(r|s)}\cdot\hat\chi_{(r'|s')}
}
\ee
With these operators, one can present (\ref{tauH}) in the form
\be
\tau^H_N=\exp\left(\sum_n \zeta_n\hat C_n\right)\cdot\sum_R\chi_R\{p_k\}\chi_R\{\bar p_k\}=
\exp\left(\sum_n \zeta_n\hat C_n\right)\cdot \exp\Big(\sum_{k=1}{p_k\bar p_k\over k}\Big)
\ee
These Casimir operators $\hat C_n$ are elements of the $GL(\infty)$ and, thus, provide the B\"acklund transformation: acting on the trivial $\tau$-function $\exp\Big(\sum_{k=1}{p_k\bar p_k\over k}\Big)$, they still give rise to a new $\tau$-function. Similarly, one can generate the KP $\tau$-function (\ref{2}) from (\ref{1}):
\be\label{87}
\tau^{(r)}=\exp\left(\sum_{i=1}^r \sum_n \zeta_n^{(i)}\hat C_n(p^{(i)})\right)\cdot
\tau_0^{(r)}
\ee
Thus, we come to the conclusion that {\bf the B\"acklund transformation is given by exponential of a linear combination of the $\hat W$-operators iff these $\hat W$-operators are from the space spanned by all single-hook operators}.

\section{Macdonald Hamiltonian and Casimir operators}

It is well-known (see, e.g., a review \cite{MMHam}) that the Macdonald polynomials form a set of
common eigenfunctions of the Ruijsenaars exponential Hamiltonians \cite{RHam}
$\hat H_m$, the simplest of which is (in fact, just this Hamiltonian is enough to fix the Macdonald polynomials unambiguously)
\be
\hat H_1 =
 \oint\frac{dz}{z}
 \exp\left(\sum_{k>0}{(1-t^{-2k})\,p_kz^k\over k}\right)\cdot
 \exp\left(\sum_{k>0}
 {q^{2k}-1\over z^k}
{\partial\over\partial p_k}\right)
\label{RHam1}
\ee
Choosing the parameter $t=q$, one returns to the Schur functions, while the system becomes the Calogero-Moser-Sutherland system at a special value of the coupling constant where the system becomes free. All higher Hamiltonians in this case are generated from this one,
\be
\hat H(q) :={1\over q-q^{-1}}
 \oint\frac{dz}{z}
 \exp\left(\sum_{k>0}{(1-q^{-2k})\,p_kz^k\over k}\right)\cdot
 \exp\left(\sum_{k>0}
 {q^{2k}-1\over z^k}
{\partial\over\partial p_k}\right)
\label{CHam}\\
\hat H(q)\chi{_R}\{p\}
= \left(\sum_{i=1}^{l_R} \frac{q^{2r_i}-1}{q^{2i-1}}+1\right)
\cdot \chi{_R}\{p\}
\label{evRMac}
\ee
since the Schur functions do not depend on $q$, and
\be
[\hat H(q),\hat H(q')]=0
\ee
Thus, $H(q)$ is a generating function of all Hamiltonians in this case. One can introduce the variable $\hbar$, $q:=e^\hbar$ and consider $\hat H(q)$ as a power series in $\hbar$. In the hook coordinates, this generating function is equal to \cite{KHam}
\be\label{99}
\hat H(q)=(q-q^{-1})\sum_{\stackrel{r,s,r',s'=1}{r+s=r'+s' }}^\infty
(-)^{s+s'}q^{r+r'-s-s'} \cdot \chi_{(r|s)}\cdot\hat\chi_{(r'|s')}=\nn\\
=2\sinh\hbar\cdot
\sum_{\stackrel{r,s,r',s'=1}{r+s=r'+s' }}^\infty
(-)^{s+s'}e^{(r+r'-s-s')\hbar} \cdot \chi_{(r|s)}\cdot\hat\chi_{(r'|s')}
\ee
Using formula (\ref{Cas}), one obtains
\be\label{gf}
\hat C(q):=\sum_{n=1}^\infty \hat C_n{(2\hbar)^n\over (n-1)!}=\hat H(q)-1
\ee
The factor $q-q^{-1}$ in (\ref{99}) explains the origin of the structure $(...+1)^n-(...-1)^n$ in the summand of (\ref{Cas}).

Thus, this generating function of the Casimir operators is exactly the Hamiltonian (\ref{CHam}), and the eigenvalue of $\hat C(q)$ is
\be\label{ev}
\hat C(q)\cdot\chi_R=\lambda(q)\cdot\chi_R,\nn\\
\lambda(q)=\sum_{i=1}^{l_R} \frac{q^{2r_i}-1}{q^{2i-1}}=\sum_i \Big(q^{2\alpha_i-1}-q^{1-2\beta_i}\Big)
\ee

\section{Conclusion}

We used the hook (Frobenius) parametrization of the Young diagrams and discussed in these terms explicit expressions for the Schur functions and shifted Schur functions, as well as for the skew Schur functions. We demonstrated that 
\begin{itemize}
\item A periodic product of the Schur functions each depending on its own set of time variables, (\ref{1}) gives rise to a KP $\tau$-function w.r.t. to all these sets of times;
\item Acting on this product with exponential of any linear combination of cut-and-join $\hat W$-operators that involve only single-hook Young diagrams is a B\"acklund transformation keeping the KP $\tau$-function (\ref{2}), (\ref{87});
\item At any level, there is exactly one cut-and-join $\hat W$-operator of this type up to a linear combination of lower level operators, sec.\ref{s7}.
\item We also explicitly constructed the simplest looking basis of these operators, the Casimir operators associated with the completed cycles (\ref{cc}), (\ref{Cas}).
\item  We evaluated the generating function of these operators (\ref{gf}) and its eigenvalue (\ref{ev}).
\end{itemize}

Remained for the future work are two big directions: $q$-$t$ generalization to the Macdonald family, and its alternative restriction to the $Q$ Schur functions \cite{QSchur,Mac,MMNs}, both require non-trivial additional ideas which can shed light on associated deformations of integrable structures.

\section*{Acknowledgements}

We are grateful to A. Orlov for the fruitful discussions. Our work is partly supported by the grant of the Foundation for the Advancement of Theoretical Physics ``BASIS" (A.Mir., A.Mor.), by  RFBR grants 19-01-00680 (A.Mir.) and 19-02-00815 (A.Mor.), by joint grants 19-51-53014-GFEN-a (A.Mir., A.Mor.), 19-51-50008-YaF-a (A.Mir.), 18-51-05015-Arm-a (A.Mir., A.Mor.), 18-51-45010-IND-a (A.Mir., A.Mor.). The work was also partly funded by RFBR and NSFB according to the research project 19-51-18006 (A.Mir., A.Mor.).  We also acknowledge the hospitality of KITP and partial support by the National Science Foundation under Grant No. NSF PHY-1748958 at the final stage of this project.

\end{document}